\documentclass[aps,prd,showpacs,nofootinbib,preprintnumbers,twocolumn]{revtex4-1}
\usepackage[ansinew]{inputenc} 
\usepackage{graphicx,epsfig}
\usepackage{url}
\usepackage{color}
\usepackage{bm}
\usepackage{dcolumn}
\usepackage{amsfonts,amssymb,amsmath}
\usepackage{psfrag}
\usepackage{subfigure}
\usepackage{latexsym}
\begin{document}
\title{Effect of the modified gravity on the large scale structure formation}
\author{Ajay Kumar  Sharma}
\email{aksh.sharma2@gmail.com}
\affiliation{Department of Physics, University of Lucknow, Lucknow, 226007, India}
\author{Murli Manohar Verma}
\email{sunilmmv@yahoo.com}
\affiliation{Department of Physics, University of Lucknow, Lucknow, 226007, India}
\date{\today}
\begin{abstract}
 We investigate  the formation  of the large scale structures in the present accelerated era  in $f(R)$ gravity background. This is done by  considering  the linear growth of matter perturbations at low redshift $z<1$. The effect of $f(R)$ alters the behaviour of the matter density perturbations from the matter dominated universe to the late-time accelerated universe which is encoded in the Newtonian gravitational constant as $G\rightarrow G_{eff}$. The modified gravitational constant ($G_{eff}$) depends on the form of $f(R)$. The late-time accelerated expansion affects the formation of large scale structures by slowing down the growth of matter density. On the other hand,  $f(R)$  increases the growth rate of the matter density perturbations. We have found that the  source term in $f(R)$ background, $G_{eff}\Omega_m$   overcomes the accelerated expansion and the effect of accelerated expansion suppresses the formation of the large scale structures in the asymptotic future.
\end{abstract}

\maketitle
\section{\label{1}Introduction}
Today we live in the phase of accelerated expansion \citep{1997ApJ...483..565P, Riess_1998} of the universe which  has the dynamical effect similar to inflation in the very beginning. We require the matter which has sufficient negative pressure to explain accelerated expansion of the universe but the known matter does not have such negative pressure. We can handle this problem by adding exotic matter in the gravitational Lagrangian density. These models are known as Dark Energy (DE) models \citep{Sahni:1999gb, Padmanabhan:2002ji, Copeland:2006wr, Sahni:2008zz, Verma:2010zza}. We can also consider the modification in the  Lagrangian density of the gravitational part of  Einstein's theory of general relativity to explain the accelerated expansion of the universe, broadly called as Modified theory of Gravity (MG).

In the present work, we  consider the $f(R)$ theory of gravity as an effective  theory and study the large scale  structure formation. Of course,  in the $f(R)$ theory of gravity, we replace Ricci scalar $R$ by a general function of $R$ in the Einstein-Hilbert action \citep{Capozziello:2002rd} and $f(R)$ reduced to the GR in the $R>>R_o = H_o^2$ regime.

The astrophysical structures existing today in the universe, had started evolving before the inflation in the very early universe. Motherland of initial seeds of the large scale structures is inflation and created due to quantum fluctuation in the inflaton field in this period. These seeds started growing after decoupling of the matter from the radiation  but in different ways in the Radiation Dominated (RD), Matter Dominated (MD) and the late-time cosmic acceleration phases. If we use DE models and $f(R)$ models to explain accelerated expansion, they provide the same results after tuning the free parameters but the growth of matter perturbations is different in both approaches \citep{Starobinsky:1998fr}.  Matter density perturbations are used as a tool to distinguish the Dark Energy (DE) models in GR background  $\Lambda$ Cold Dark Matter ($\Lambda$CDM) from $f(R)$ gravity models \citep{Starobinsky:1998fr, Narikawa:2009ux, Gannouji:2008wt}. The matter density perturbations depend on the models of $f(R)$ gravity, therefore their behaviour varying with the parameters of the  models.

Some of the most popular models in $f(R)$ gravity are \citep{Starobinsky:2007hu, Hu:2007nk, Percival:2007yw, Appleby:2007vb, Amendola:2006we, Tsujikawa:2007xu} and Dvali-Gabadadze-Porrati (DGP) brane-world model \citep{Dvali:2000hr} beyond the $\Lambda$CDM model. In this paper we have used the Power-law models \citep{Sharma:2022tce, Capozziello:2003tk, Carloni:2004kp, PhysRevD.91.064016, Jaime:2012yi, KumarSharma:2022qdf, Sharma:2020vex, Sharma:2019yix, Yadav:2018llv}, Starobinsky model \citep{Starobinsky:1998fr}, and $\Lambda$CDM model \citep{Sahni:1999gb, Padmanabhan:2002ji} to study the matter density perturbations and compare their results with observations \citep{2dFGRS:2005yhx} in the linear regime.

This paper is organized in five sections. In Section II we give the basic introduction of the modified theory of gravity and its dynamical equations. We discuss  the cosmological  perturbations in the linear regime  and matter density perturbations in Section III. In Section IV, we investigate the behaviour of the growth index and modified gravitational constant. We also discuss the nature of matter density perturbations in the asymptotic future. At the end, in Section V we give a summary of the results of our work.

(Throughout the paper, we have used the $c = 1$ and $\kappa^2 = 8\pi G$.)

\section{\label{2} Modified  $f(R) $ background }
A simple way to modify the gravity is to replace the Ricci curvature $R$ by a function of $R$  in the Einstein-Hilbert action \citep{Sotiriou:2008rp,DeFelice:2010aj, amendola_tsujikawa_2010, Fazio:2018djb}. Thus, the  action in $f(R)$ gravity is given as
\begin{eqnarray} \mathcal{A} = \frac{1}{2\kappa^2} \int{d^4x\sqrt{-g}  f(R)} \nonumber \\ +\int{d^4x\sqrt{-g}  \mathcal{L}_m}(g_{\mu\nu}, \Psi_m), \label{IIA1}\end{eqnarray}
where $\mathcal{L}_m$ is the matter field Lagrangian density minimally coupled with curvature. Taking the variation of \eqref{IIA1} with respect to metric tensor $g_{\mu\nu}$,  we obtain the gravitational field equations as
\begin{eqnarray}F(R)R_{\mu\nu} - \dfrac{1}{2}f(R)g_{\mu\nu}- \nabla_\mu\nabla_\nu F(R)\nonumber \\+ g_{\mu\nu}\square F(R)= \kappa^2 T_{\mu\nu}, \label{IIA2}\end{eqnarray}
where  $  \square \equiv g^{\mu\nu} \nabla_\mu\nabla_\nu$ is the  covariant d'Alembertian operator, $F(R) = {\partial f}/{\partial R}$, $\nabla_\mu$ is the  covariant derivative, $T_{\mu\nu}$ is the energy-momentum tensor of matter filed and it should satisfy the energy-momentum conservation equation $ \nabla^\mu T^{(m)}_{\mu\nu} = 0$ of the matter field. Trace of the  equation \eqref{IIA2} is
\begin{eqnarray} 3\square F(R) + F(R)R - 2f(R)= \kappa^2 T, \label{IIA4}\end{eqnarray}
where $T = g_{\mu\nu}T_{(m)}^{\mu\nu}= -\rho_m +3 p_m$ and the suffix $m$ denotes pressureless matter (including dark matter).  Ricci scalar in the flat FLRW background metric is, $R = 6(\dot H +2H^2)$. First derivative of the $f(R)$ gives a new scalar degree of freedom which denotes the deviation from General Relativity (GR). If we consider the de Sitter universe, then the Hubble expansion rate is constant $(\dot H = 0)$. Therefore, the Ricci scalar becames constant and  $\square F = 0$ at the de-Sitter  point. We can rewrite the equation \eqref{IIA4} for de-Sitter point as
\begin{eqnarray} F(R)R - 2f(R)= \kappa^2 T, \label{IIA5}\end{eqnarray}
and the solution of the above equation \eqref{IIA5} becomes
\begin{eqnarray} R_{min} = \frac{2f(R_{min}) + \kappa^2 T}{F(R_{min})}, \label{IIA6}\end{eqnarray}
where $R_{min}$ is the value of Ricci scalar at de-Sitter point.

Equation  \eqref{IIA2} can  also be written as
\begin{eqnarray} G_{\mu\nu}\equiv R_{\mu\nu} -\frac{1}{2}Rg_{\mu\nu} = 8\pi G\Big[T_{\mu\nu} + \nonumber \\  g_{\mu\nu} \Big( \dfrac{f(R) - RF}{2}\Big) + (\nabla_\mu\nabla_\nu F(R) -g_{\mu\nu}\square F(R))\Big]. \label{IIA7} \end{eqnarray}
 We can, in a straightforward way, compare the above equation \eqref{IIA7} with the Einstein's equation $ R_{\mu\nu} -(1/2)R g_{\mu\nu} = \kappa^2 T_{\mu\nu}$ in GR. We have modified energy-momentum tensor as $ T_{(m)}^{\mu\nu} + T_{f(R)}^{\mu\nu}$.

We obtain Friedmann equations in  the $f(R)$ gravity as \citep{Huang:2014fua}
\begin{eqnarray} 3H^2 = 8\pi G\rho_m + \Big[ \frac{RF -f(R}{2} - 3H\dot F -3H^2 (F-1)\Big],  \label{IIA8}\end{eqnarray}
\begin{eqnarray} -2\dot H = 8\pi G\rho_m + \Big[ \ddot F - H\dot F + 2\dot H(F-1) \Big] , \label{IIA9} \end{eqnarray}
where  $\ddot a/a = \dot H +H^2$ and $\dot a/a \equiv H$.
Some extra terms appear in the Friedmann equations \eqref{IIA8} and  \eqref{IIA9} due to $f(R)$ gravity, which  act as pressure and energy density due to $f(R)$ gravity.
We have an effective equation of state in the $f(R)$ gravity given as
\begin{eqnarray}w_{eff} = -1 -\frac{2\dot H}{3H^2}. \label{IIA10}\end{eqnarray}
We  obtain the effect of acceleration in the $f(R)$ gravity without introducing exotic form of matter as a Dark Energy (DE) component. Equation of state to explain the effect of the late-time acceleration is  $w = -1.03\pm0.03 $ in \citep{Planck:2018vyg}. The expression of $w_{eff}$ \citep{Sharma:2022tce} for model $f(R)= R +R^{1+\delta}/R_c^\delta$  is
\begin{eqnarray} w_{eff} = -1+ \frac{2}{3}\varepsilon_1  = -1 + \frac{2(1-\delta)}{3\delta(1+2\delta)}. \label{IIA11} \end{eqnarray}
where $\varepsilon_1 \simeq (1-\delta)/[3\delta(1+2\delta)]$ and $\delta$ is a model parameter.
\begin{figure}[h]
\centering  \begin{center} \end{center}
\includegraphics[width=0.50\textwidth,origin=c,angle=0]{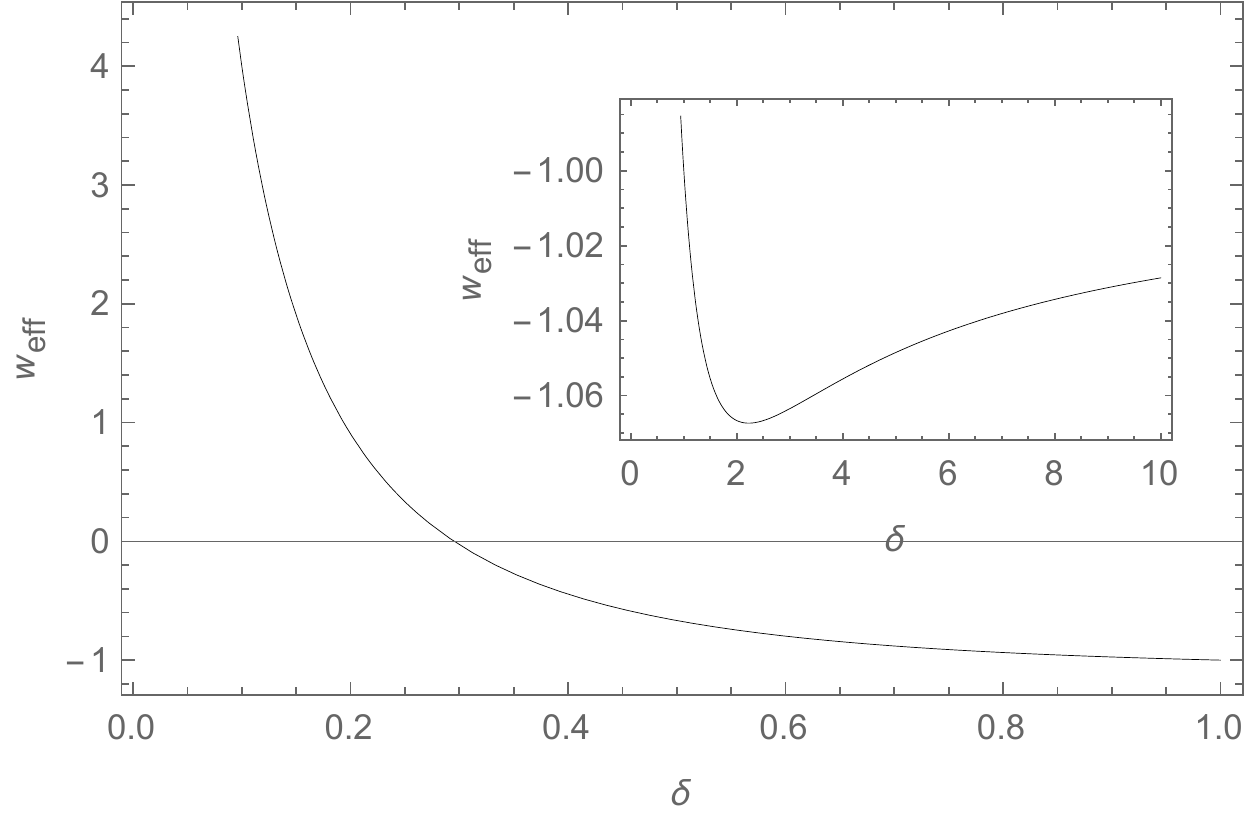}
\caption{\label{fig:p2} Plot between the $w_{eff} $ and model parameter $\delta$. We have found that $w_{eff} $ gains both positive and negative values at different $\delta$. There is a boundary line at $\delta \simeq 0.3$. } \label{EoS}
\end{figure}
The value of $w_{eff} \simeq -0.998$ at $\delta = 0.98$ which is consistent with Planck 2018  data \citep{Planck:2018vyg}. Figure \eqref{EoS} clearly shows that the value of $w_{eff}$ never goes below $-1.02$ at $\delta = 2.2247 $. This value is inside  the present observational limits.   We can recover the $\Lambda$CDM model  $w_{eff} =-1$ at  $\delta = 1$.  The quantitative behaviour of the model parameter $\delta$ for acceleration and inflation has been discussed in detail in our previous work  \citep{Sharma:2022tce}. We obtained the range of $\delta$ which is $0.366<\delta<1$ for acceleration and for inflation (i.e. for quasi-de Sitter universe) $\delta\approx 0.98$. Therefore, we restrict to  this limit and use $\delta=0.98$ for late-time cosmic acceleration as well.

\section{\label{4} Matter density Perturbations in $f(R)$ gravity }
Now, we study matter density perturbations in the linear regime which enables us to distinguish dark energy models in $f(R)$ gravity from the other theories of Dark Energy in GR. We can also distinguish $\Lambda$CDM model from $f(R)$ theories.  We take a  metric perturbation about the flat FLRW background \citep{DeFelice:2010aj}
\begin{eqnarray}ds^2 = -(1 +2\alpha)dt^2 -2a(t)[\partial_i\beta -S_i]dtdx^i \nonumber \\+a^2(t)[\delta_{ij} +2\psi\delta_{ij}  +2\partial_i\partial_j\gamma +2 \partial_jF_i +h_{ij} ]dx^idx^j, \label{IIIA1}\end{eqnarray}
where $\alpha$, $\beta$, $\gamma$, $\psi$ are scalar perturbations, $S_i$, $F_i$ are vector perturbations, and $h_{ij}$ is the tensor perturbation. In this paper we focus on the scalar perturbations only. After using the Longitudinal gauge, the line element without vector and tensor perturbations is given  \citep{DeFelice:2010aj} as
\begin{eqnarray}ds^2 = -(1 +2\Phi)dt^2  +a^2(t)(1-2\Psi )\delta_{ij}dx^idx^j. \label{IIIA2} \end{eqnarray}
We include perturbations in the non-relativistic matter with negligible pressure, $p_m = 0$ in $f(R)$ background  \citep{DeFelice:2010aj}
\begin{eqnarray}\delta\dot\rho_m +3H\delta\rho_m = \rho_m\Big( A -3H\alpha - \frac{k^2}{a^2}v\Big), \label{IIIA3} \end{eqnarray}
where $A\equiv 3(H\alpha - \dot\psi) + \frac{k^2}{a^2}\chi$, $\chi\equiv a(\beta +a \dot\gamma)$. After taking the  longitudinal gauge  $\alpha = \Phi$, $\psi = \Psi$ and  consider  quasi-static approximation in $f(R)$ background \cite{Boisseau:2000pr, Copeland:2006wr, Tsujikawa:2007gd, Tsujikawa:2007tg}, when wave number $k$ keeps deep inside the Hubble radius $(k>>aH)$
\begin{eqnarray} \Big[\frac{k^2}{a^2}|\Phi|, \frac{k^2}{a^2}|\Psi|, \frac{k^2}{a^2}|\delta F|, M^2|\delta F|  \Big] \nonumber \\ \gg ( H^2|\Phi|,  H^2|\Psi|,  H^2|B|, H^2|\delta F|),          \label{SNIIIA3.3} \end{eqnarray}
and
\begin{eqnarray} |\dot X| \lesssim |HX|,      \label{SNIIIA3.4} \end{eqnarray}
where $\dot X = \Phi, \Psi, F, \dot F, \delta F, \dot{\delta F} $.
Thus the equation for matter density perturbations becomes  as \citep{DeFelice:2010aj}
\begin{eqnarray}\ddot\delta_m +2H\dot\delta_m -4\pi G_{eff} \rho_m\delta_m \simeq 0, \label{IIIA4} \end{eqnarray}
where $\delta_m = \delta\rho_m/\rho_m$ and $G_{eff}$ is the effective gravitational coupling given in \citep{Tsujikawa:2007gd, Tsujikawa:2007tg} as
\begin{eqnarray}G_{eff} = \frac{G}{F}\frac{4k^2/a^2 + 3 M^2}{3(k^2/a^2+ M^2)}, \label{IIIA5}\end{eqnarray}
where $M^2$ is the mass of the new degree of freedom given as
\begin{eqnarray}M^2 = \frac{1}{3}\Big[ \frac{F(R_{min})}{R_{min}F_R(R_{min})}  - R_{min}\Big], \label{IIIA5}\end{eqnarray}
with $F_R \equiv \partial^2 f(R)/{\partial R}^2$. In the equation \eqref{IIIA4}, the second term has the Hubble parameter. It carries the information of expansion of the universe and $G_{eff} $  contains the information about the modification of the gravity through $f(R)$. In the high density region ($R_{min}>> R$)  equation \eqref{IIIA5} becomes $M^2 \simeq 1/3F_R$ \citep{Tsujikawa:2007tg}. For viable $f(R)$ gravity models, we have assumed $|F(R)| << 1$  and $|F_{R}| << 1$. Then $G_{eff}/G$ can be further written in the \citep{Narikawa:2009ux, Gannouji:2008wt} simplified way as
\begin{eqnarray}\frac{G_{eff}}{G} \simeq 1 + \frac{1}{3}\frac{k^2/a^2}{(a^2/k^2 + 1/3f_{RR})}, \label{IIIA6}\end{eqnarray}
where $f_{RR}\equiv \partial^2 f(R)/{\partial R}^2$. Above relation shows that the $G_{eff}$  depends on the wave number $k$ and on the $f(R)$ model parameter. In the regime where $ k/a << M $ the effective gravitational constant becomes $G_{eff} \simeq G$, and mimics GR.  In the scalar-tensor regime  $ k/a >> M $, $G_{eff}$ becomes $\simeq 4G/3$, from equation \eqref{IIIA6}. The transition from GR to scalar-tensor regime occurs during the matter dominated phase at  $ k/a = M $.

Now, the equation \eqref{IIIA4}, after changing the $d/dt\rightarrow d/dN$ and a growth parameter $f \equiv d\ln \delta_m/dN= \Omega_m^\gamma $ (where $\gamma$ is the growth index) becomes as
\begin{eqnarray}\frac{df}{d\ln a} + f^2 + \Big(2+  \frac{\dot H}{H^2}\Big)f  = \frac{3}{2}\Omega_m \frac{G_{eff}}{G} , \label{IIIA8}\end{eqnarray}
where $N \equiv \ln{a}$  and ${\Omega}_m = 8\pi G \rho_m/(3H^2)$.
We have found the constant growing mode of the matter density perturbation deep inside the matter dominated universe from the equation \eqref{IIIA8} if  $f = p$ in the \citep{Gannouji:2008wt, Gannouji:2008jr} is
\begin{eqnarray}\delta_m = A a^{P_1} +B a^{P_2} , \label{IIIA13}\end{eqnarray}
where $P_1 = (1/4)(-1 + \sqrt{1+24C})$, $P_2 = (1/4)(-1 -  \sqrt{1+24C})$ and $C \equiv (G_{eff}/G) \Omega_m =$ constant. This solution is also valid for $z<1 $.  In the regime where $k/a<<M$ we have $C \approx 1$ and $C \approx (4/3)$ for $k/a>>M$. We get $\delta_m$ in \citep{Starobinsky:2007hu} as
\begin{align}
\delta_m \propto a \propto t^{2/3} && \text{if} &&  k/a << M,  \nonumber \\
\delta_m \propto a^{\frac{\sqrt{33}-1}{4}} \propto t^{\frac{\sqrt{33}-1}{6}} && \text{if} &&  k/a >> M.
\label{IIIA14}\end{align}

If we consider the growth index $\gamma$, as a function of redshift $z$, the equation \eqref{IIIA8} can be expressed in terms of $\Omega_m$, $w_{DE}$, $\gamma$ and $d\gamma/dz = \gamma\prime$, using $ f = \Omega_m^{\gamma(z)}$ as
\begin{eqnarray}-(1+z)(\ln \Omega_m)\gamma^\prime +\Omega_m^\gamma + \frac{1}{2} \Big(1+  3(2\gamma -1) w_{eff}\Big) \nonumber \\=  \frac{3}{2} \frac{G_{eff}}{G}\Omega_m^{1-\gamma} , \label{IIIA9}\end{eqnarray}
and $\gamma\prime_o$ at redshift $z =0$ is
\begin{eqnarray}\gamma_o^\prime = \frac{ \Omega_{mo}^{\gamma_o} +  3(\gamma_o -0.5) w_{eff} - \frac{3}{2} \frac{G_{eff}}{G_{N}}\Omega_{mo}^{1-\gamma_o}- \frac{1}{2}}{\ln{\Omega_{mo}}}, \label{IIIA10}\end{eqnarray}
where $\Omega_m$ is the matter density parameter and $o$ indicates the present values of the parameters.
 $G_{eff}/G = 1$ in the $\Lambda$CDM model in GR background and the  above expression becomes
\begin{eqnarray}\gamma_o^\prime = \frac{ \Omega_{mo}^{\gamma_o} +  3(\gamma_o -0.5) w_{eff} - \frac{3}{2} \Omega_{mo}^{1-\gamma_o}- \frac{1}{2}}{\ln{\Omega_{mo}}}. \label{IIIA11}\end{eqnarray}

We have found the constraint on the growth index as $f(\gamma_o, \gamma^\prime_o, \Omega_{mo}, w_{eff})=0$ and growth index in the DE models in GR given in \citep{Wang:1998gt, Nesseris:2007pa} as
\begin{eqnarray}\gamma = \frac{3(w_{eff}-1)}{ (6w_{eff} -5)} +\frac{3}{125} \frac{(1-w_{eff})(1-3w_{eff}/2)}{(1-6/5 w_{eff})^3}(1-\Omega_m), \label{IIIA12}\end{eqnarray}
 where $\gamma $ takes constant value $\sim 0.56$. However, we considered  dependence of $\gamma$ on the redshift $z$ beyond the GR. Therefore we can expand  it around $z$ as $ \gamma(z) \simeq \gamma_o + \gamma^\prime_o z$ for $0\leq z \leq 0.5$ in the \citep{Polarski:2007rr}.

The growth of the matter perturbation changes after entering into the scalar-tensor regime till the time $t_{acc}$, which is the epoch at which universe enters the late-time accelerated phase.

\section{\label{4} Model dependence on the  growth of matter perturbations }
We can distinguish  the dark energy models in GR background and beyond the GR on the basis of power spectrum of the growth of matter perturbations $\Delta n_s$, growth index $\gamma$,  growth factor $f$ and $G_{eff}$.

\subsection{\label{4.1} $\Lambda$CDM model}
In the $\Lambda $CDM model, $\Lambda$ term is used to explain the late-time cosmic acceleration with cold dark matter. In this model, the equation of state is constant $w_{\Lambda} = -1$ and $G_{eff} = G_N$ because $df/dR = 1$.  We have $\Delta n_s = 0$ here, and the non-zero $\Delta n_s$ indicates a theory beyond the GR.

In the previous works \citep{Peebles:1984ge, Lahav:1991wc}, several authors obtained approximately constant growth index $\gamma$ with negligible dependence on the $\Omega_{m}$. Some papers \citep{Wang:1998gt, Gannouji:2008jr, Gannouji:2008wt} show $\gamma_o =6/11 \approx 0.5454$.

However, some authors realised that the growth index might be a function of redshift and therefore, the deviation from constant $\gamma$ provides a new constraint on the theories beyond GR. This information is stored in the $\gamma^\prime$.  For $\Lambda$CDM, \citep{ Gannouji:2008jr} growth index and its derivative are $  0.554\leqslant\gamma_o\leqslant 0.558$ and $ -0.0195\leqslant\gamma^\prime_o\leqslant- 0.0157$ for $0.2\leqslant\Omega_{mo} \leqslant 0.35$. Many other models in GR background \citep{ Gannouji:2008jr}  have $|d\gamma/dz \equiv \gamma^\prime_o(z=0)|\lesssim 0.02$ with slight dependence on $\Omega_{mo}$. But for slowly varying $w_{DE}$, and $\Omega_m = 0.3$ the growth index is slightly higher $\gamma_o = 0.555$ in \citep{Gannouji:2008wt}. Higher and positive value of the $\gamma^\prime_o$ is clear signal of the departure from the $\Lambda$CDM model. Authors have found the bound on the $\Delta n_s$ conservatively in \citep{Gannouji:2008wt} as
\begin{eqnarray} n_s^{gal} - n_s^{CMB} =\Delta n_s < 0.05, \label{IVB2} \end{eqnarray}
 However, the observations \citep{Tegmark:2006az} do not allow for any significant difference in between the slopes of the two power spectra.

\subsection{\label{4.2} Starobinsky model}

In this part of the paper, we have used the Starobinsky model in the $f(R)$  gravity beyond the GR background, gives as
\begin{eqnarray} f(R) = R - \lambda R_c \Big[ 1- \Big( 1+ \frac{R^2}{R_c^2}\Big)^{-n}\Big] + \alpha R^2, \label{IVB1}\end{eqnarray}
where $\lambda $, $n >0$, and $\beta R_c \simeq 2\Lambda(\infty)$ \citep{Gannouji:2008wt}. In the above model, second term is used to explain the late-time cosmic acceleration and third term for inflation (very early acceleration).

In the Starobinsky model, one finds \citep{Gannouji:2008wt} a distinction from the $\Lambda$CDM model on basis of the evolution of matter perturbations. There appears a difference in the spectral index $n_s^{CMB}$ of primordial power spectrum $P(k) \propto k^{n_s^{CMB}}$ and spectral index $n_s^{gal}$, calculated from the galaxy power spectrum $P(k) \propto k^{n_s^{gal}}$.
We  obtained $\Delta n_s$ in the \citep{Starobinsky:2007hu} after dropping the $\alpha R^2$  from the  model given by \eqref{IVB1}.  In the matter dominated era at  $k/a << M$
\begin{align}
 a \propto t^{2/3} &&  R_{o} = \frac{4}{3t }; && f_{,RR} \propto t^{4n+4}; && M(R) \propto t^{-2n-2},
\label{IVB3}\end{align}
and $\delta_m$ grows as $ t^{2/3}$ and it changes after $k = aM(R)$ as $t^{(\sqrt{33}-1)/6} $ and to  $t_{acc}$,  the  end of matter dominated epoch \citep{Starobinsky:2007hu, DeFelice:2010aj}. Thus
\begin{eqnarray} \frac{P_{\delta_m}(t_{acc})}{P_{\delta_m}^{\Lambda CDM}(t_{acc})} =  \Big(\frac{t_{acc}}{t_k}\Big)^{2\Big(\frac{\sqrt{33}-1}{6}- \frac{2}{3}\Big)} \propto k^{\frac{\sqrt{33}-5}{6n+4}}, \label{IVB4}\end{eqnarray}
where time $t=t_{acc}$ corresponds to the end of the matter dominated phase and at that epoch the universe enters into the later-time acceleration phase.
Now, we have $\Delta n_s$  \citep{Starobinsky:2007hu, DeFelice:2010aj}
\begin{eqnarray} n_s^{gal} - n_s^{CMB} = \frac{\sqrt{33}-5}{2(3n+2)}. \label{IVB5} \end{eqnarray}
The above expression depends on the model parameter $n$. If $n$ increases then $\Delta n_s$ shifts towards the \citep{Tegmark:2006az} bound given by the equation \eqref{IVB2}. Therefore, the constraint on $n$ is $n\geq 2$. For $n =1$, $\Delta n_s = 0.074$ and  $\Delta n_s = 0.047$ for $n=2$ at $k = 3000 a_o H_o$, $z_{acc} = 0.7$ and $z_k(t_k)= 2.5$.

In this model, the growth index $\gamma_o(z=0) \simeq 0.41$  is smaller than the $\Lambda$CDM model in GR background. We have also find out the $\gamma_o \simeq 0.4$, $\gamma^\prime_o\simeq -0.25$ for $\Omega_{mo}= 0.32$ and  $\gamma_o \simeq 0.43$, $\gamma^\prime_o\simeq -0.18$ for $\Omega_{mo}= 0.23$ in  \citep{Gannouji:2008jr, Gannouji:2008wt}. The matter density of the universe is lower when $\gamma_o$ and the value of $ \gamma_o^\prime$ move closer to  predicted value even though still remaining far from it.

\begin{figure}[h]
\centering  \begin{center} \end{center}
\includegraphics[width=0.50\textwidth,origin=c,angle=0]{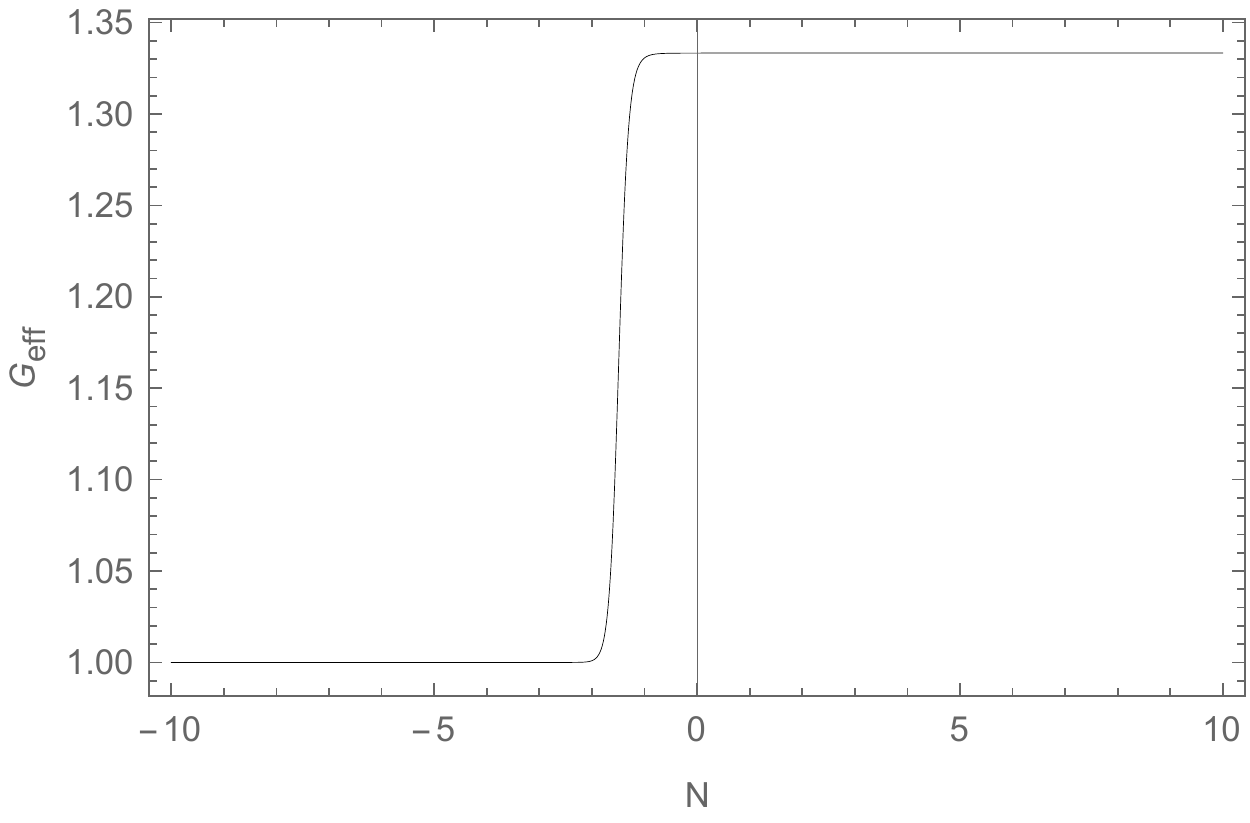}
\caption{  This is the plot between $G_{eff}$ and number of e-foldings $N$ in the Starobinsky model. We take the value of model parameters as $n =1$ and $ \lambda= 1$. It is seen that the growth never ends in this model in the asymptotic future.} \label{4.2f1}
\end{figure}

We have found the significant effect of the $f(R)$ model on the source term as
 \begin{eqnarray} \frac{G_{eff}}{G_N} \Omega_m \propto a^{3(2n+ 2) + 3w -2}, \label{VA1} \end{eqnarray}
\begin{align}
 \frac{G_{eff}}{G_N} \textcolor[rgb]{1.00, 0.00, 0.00}{\Omega_m}\propto a; &&  if && n = 0, \nonumber \\
 \frac{G_{eff}}{G_N} \textcolor[rgb]{1.00, 0.00, 0.00}{\Omega_m}\propto a^{7}; &&  if && n = 1.
\label{IVA3}\end{align}
We have used different values of $n$ at which we consider the de Sitter type background expansion i.e. $w = -1$ and  $\dot H/ H^2 = 0$.

We have found that the source terms always dominates over the  expansion term. Thus, $G_{eff}/G_N$  remains $4/3$ in the asymptotic future  and growth of the matter perturbations  never ends. We can also see it from the Fig. \eqref{4.2f1} where growth in matter density is large at the present epoch but far back in the matter dominated phase, growth rate was smaller than the present rate.

\subsection{\label{4.3} $f(R) = R + R^{1+\delta}/R_c^\delta$  model}
We aim  to study the evolution of the matter density perturbations in the our model given by the equation \eqref{IVC1} and compare it with $\Lambda$CDM model in GR background and the Starobinsky model in $f(R)$ background. Certainly, the evolution of the matter density perturbations $ \delta_m$ or  scalar perturbations begins at the pre-inflationary universe and further enters into the  radiation and matter dominated eras. Indeed, at the end, these perturbations enter into the late-time comic acceleration phase. These perturbations got freezed during inflation and started growing after re-entering the Hubble horizon. Growth of the matter density perturbation far inside the matter dominated  era is $\delta_m \propto t^{2/3}$ and after transition from GR to scalar-tensor regime at $k = aH = M(R)a$, $\delta_m $ evolves differently which depends on the model parameters. Our $f(R)$ model is given as
\begin{eqnarray} f(R) = R + \frac{R^{1+\delta}}{R_c^\delta}, \label{IVC1} \end{eqnarray}
Now we calculate scale factor $a$, $f(R)_{,RR}$ and $M^2$ from the \eqref{IVC1}  as
\begin{align}
 a \propto t^{2/3}; && f(R)_{,RR} = \frac{\delta(1+\delta)}{R_c^{\delta}}\Big(\frac{4}{3} \Big)^{\delta-1} t^{-2(\delta-1)},
\label{IVC2}\end{align}
\begin{align} M^2(R)\simeq \frac{1}{3f^{\prime\prime}} = \frac{R_c^{\delta}}{3\delta(1+\delta)}\Big(\frac{3}{4} \Big)^{\delta-1} t^{2(\delta-1)} \propto  t^{2(\delta-1)}.
\label{IVC3}\end{align}
Further using \eqref{IVC3}  to obtain the ratio ${P_{\delta_m}(t_{acc})}/{P_{\delta_m}^{\Lambda CDM}(t_{acc})}$ we have
\begin{eqnarray} \frac{P_{\delta_m}(t_{acc})}{P_{\delta_m}^{\Lambda CDM}(t_{acc})} =  \Big(\frac{t_{acc}}{t_k}\Big)^{2\Big(\frac{\sqrt{33}-1}{6}- \frac{2}{3}\Big)} \propto k^{\frac{\sqrt{33}-5}{1-3\delta}}, \label{IVC4}\end{eqnarray}
Thus, we can obtain the $\Delta n_s $ from the equation \eqref{IVC4} by using the definition $dP/d\ln{k}\equiv n_s$ \citep{amendola_tsujikawa_2010} as
\begin{eqnarray} n_s^{gal} - n_s^{CBM} = \frac{\sqrt{33}-5}{(1-3\delta)}. \label{IVC5} \end{eqnarray}

\begin{figure}[h]
\centering  \begin{center} \end{center}
\includegraphics[width=0.50\textwidth,origin=c,angle=0]{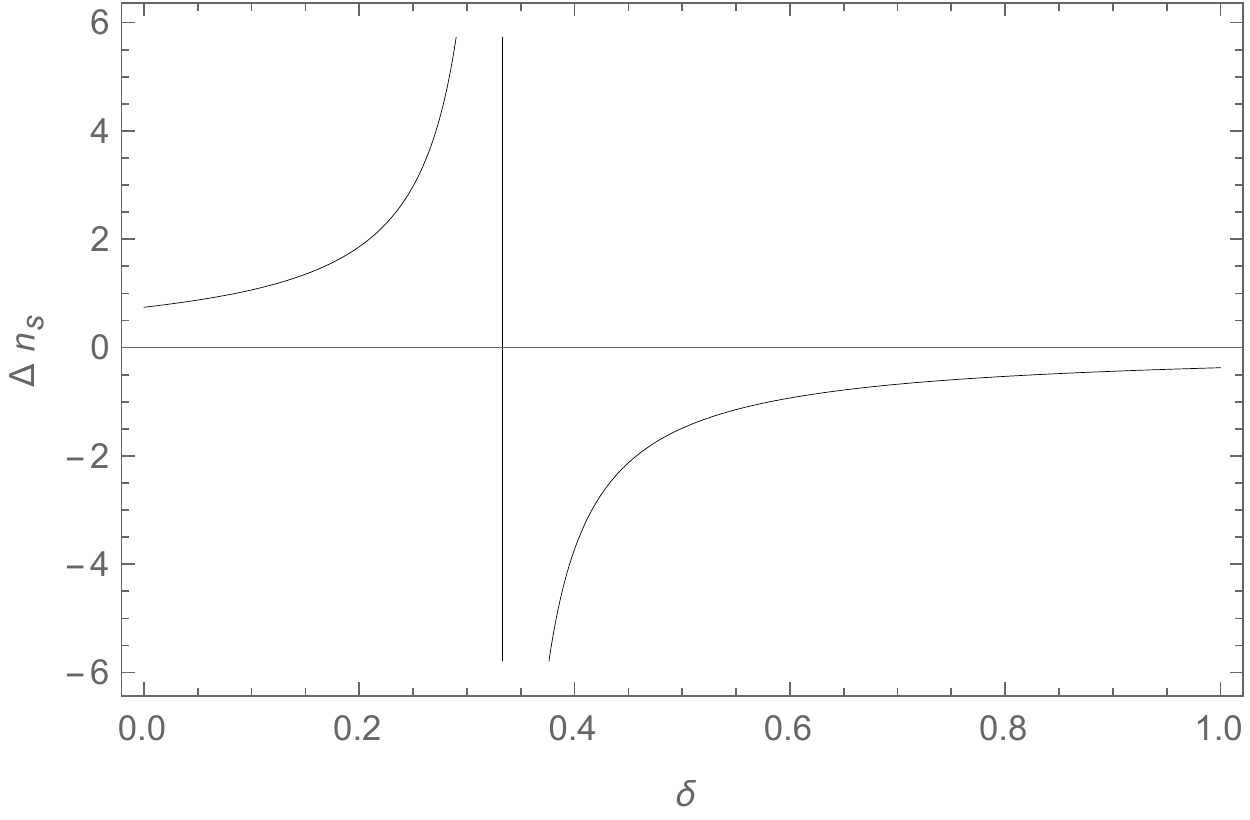}
\caption{ Plot in the left panel shows the behaviour of $\Delta n_s$ for $1-3\delta >0$ and in the right panel for $1-3\delta >0 $. For accelerated expansion we have $0.336< \delta < 1 $. This means $\Delta n_s$ is negative value, but as $\delta\rightarrow 0$,  then $\Delta n_s$ also tends to zero. That is  consistent with the current observations.}   \label{4.3f1}
\end{figure}
Equation \eqref{IVC5} depends only  on the model parameter $\delta$. Figure \eqref{4.3f1} shows that when $\delta\leq1$ then $\Delta n_s <0$,  meaning $n_s^{gal} < n_s^{CBM} $ and if $1<3\delta$, or $1/3<\delta$ then $\Delta n_s >0$. The value of $\Delta n_s \simeq -0.383795$ at $\delta = 0.98$ which is much larger than the lower bounds \citep{Tegmark:2006az} given by the equation \eqref{IVB2}.

We have found a discrepancy between the spectral index obtained from the galaxy survey $n_s^{gal}$ and primordial power spectrum from the CMB, $n_s^{CMB}$ beyond the GR background.

We have used the growth parameter $f$, growth index $\gamma$ and variation of the $\gamma$ with redshift ($\gamma^\prime(z)$) to distinguish $f(R)$ models from $\Lambda$CDM model. Information about the modification in GR is stored in the $G_{eff}$ and equation of state $w_{eff}$ in the equation \eqref{IIIA5} and \eqref{IIA10}, respectively.
At large redshift $G_{eff}\rightarrow G/F \simeq 1$ and at low redshift $G_{eff}\rightarrow 4G/3F \simeq 1.33$. $G_{eff}$ in the model given by the equation \eqref{IVC1} is
\begin{eqnarray} \frac{G_{eff}}{G_N} = 1+ \frac{1}{3}\frac{(1+z)^2(k/a_oH_o)^2H_o^2}{(1+z)^2(k/a_oH_o)^2H_o^2 + \frac{R_c^{\delta}}{3\delta(1+\delta)R^{\delta -1}}}. \end{eqnarray}
Now, we use $R_c \simeq \Lambda \simeq H_o^2$ and $R \simeq 3H_o^2[\Omega_{mo}a^{-3} +(1-3w_{eff}) \Omega_{DEo}a^{-3(1+w_{eff})}]$ in the above equation to obtain
\begin{eqnarray} \frac{G_{eff}}{G_N} = 1+\frac{1}{3} \frac{K^2a^{-2}}{K^2a^{-2} + \frac{[\Omega_{mo} a^{-3} +(1-3w_{eff}) \Omega_{DEo}a^{-3(1+w_{eff})}]^{1-\delta}}{3^\delta\delta(1+\delta)}}, \label{IVA1} \end{eqnarray}
where $K \equiv k/(a_oH_o)$ and $1+z = 1/a$.

The value of $G_{eff}$ does not seem to reach unity in the large redshift (during matter dominated era) regime. This means that the Newtonian gravitational constant is always modified in this model.

\begin{figure}[h]
\centering  \begin{center} \end{center}
\includegraphics[width=0.50\textwidth,origin=c,angle=0]{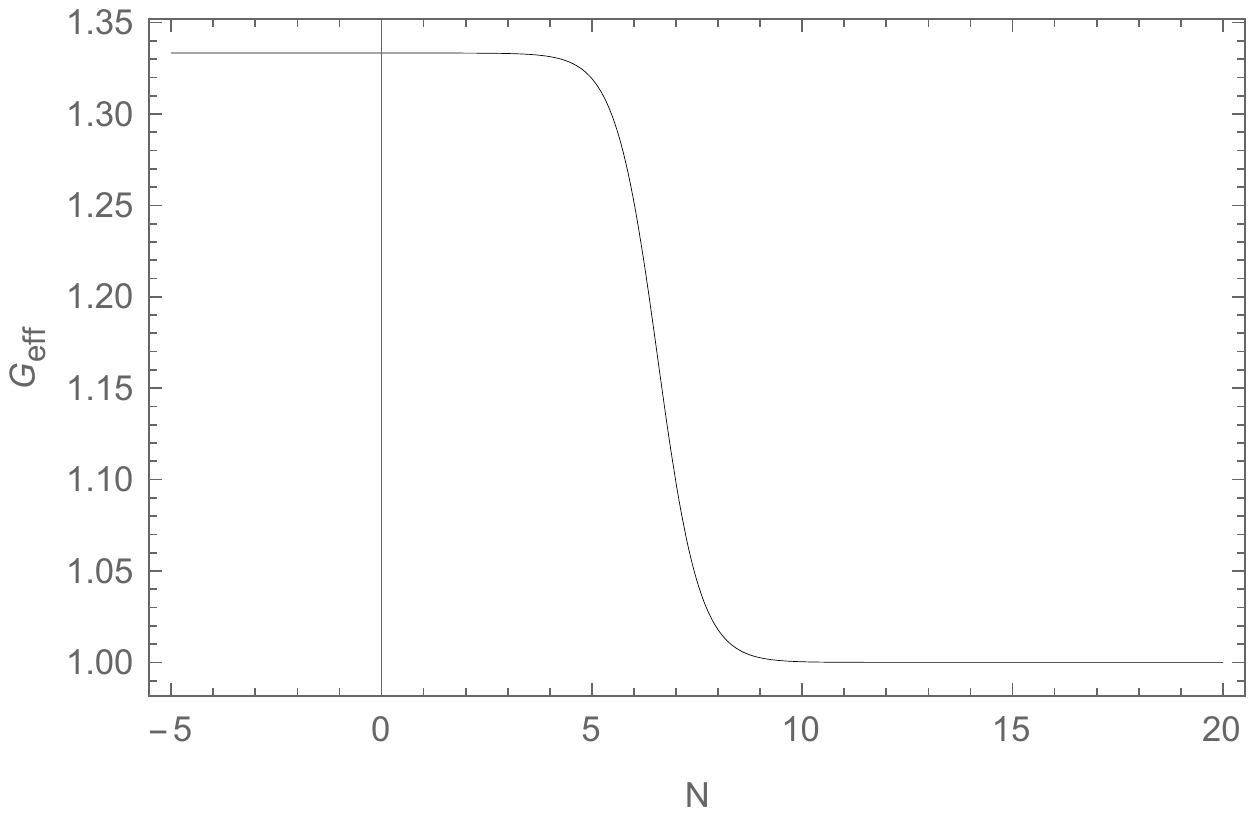}
\caption{ Plot between $G_{eff}$ and the number of e-foldings $N$ in the model given by \eqref{IVC1}.}   \label{4.3f5}
\end{figure}

Further, we focus on  the source term (last term) and expansion term (third term) or growth parameter at present and asymptotic  future. We have obtained the modified gravitational constant from the equation \eqref{IVA1} and matter density parameter,  $G_{eff}/G_N \propto a^{3(1-\delta) -2}$ and $\Omega_m \propto a^{3w_{eff}}$ \citep{Linder:2018pth} respectively. Therefore, the  combined expression becomes
 \begin{eqnarray} \frac{G_{eff}}{G_N} \Omega_m \propto a^{3(1-\delta) + 3w_{eff} -2}, \label{IVA2}\end{eqnarray}
The value $w_{eff} $ in the \eqref{IVC1} are $ \sim -0.988$ and $ \sim -1.02$ \citep{Sharma:2022tce}. Thus, we consider the equation of state for background accelerated expansion is $w_{eff} = -1$ in the equation \eqref{IVA2} leading to
\begin{align}
 \frac{G_{eff}}{G_N} \textcolor[rgb]{1.00, 0.00, 0.00}{\Omega_m}\propto a^{-2}; &&  if && \delta = 0 \nonumber \\
 \frac{G_{eff}}{G_N} \textcolor[rgb]{1.00, 0.00, 0.00}{\Omega_m}\propto a^{-5}; &&  if && \delta = 1.
\label{IVA3}\end{align}
Here we notice that $(G_{eff}/G_N) \Omega_m $ decreases in asymptotic future and  the expansion term dominates over it and growth rate of the matter perturbation ends after $\sim$5e-folds. However, at the present accelerated epoch, source term dominates over the expansion term. We can see it from the Fig.\eqref{4.3f5}.
\begin{figure}[h]
\centering  \begin{center} \end{center}
\includegraphics[width=0.50\textwidth,origin=c,angle=0]{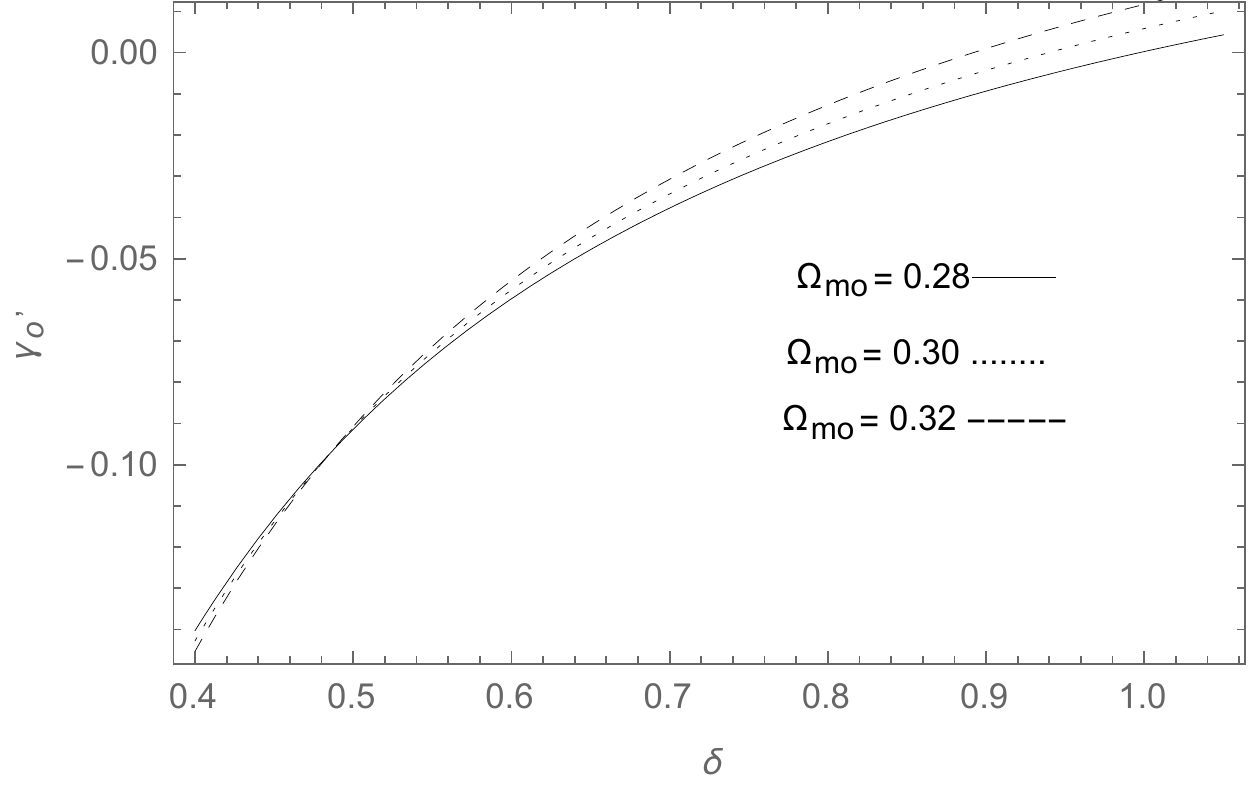}
\caption{ Plot of  $\gamma^\prime_o$ with respect to the model parameter $\delta$ for different $\Omega_{mo}$. It shows small dependence on the  $\Omega_{mo}$. It also provides a  wide range of $\delta $ for matter density perturbations in the linear regime.} \label{4.3f3}
\end{figure}

\begin{figure}[h]
\centering  \begin{center} \end{center}
\includegraphics[width=0.50\textwidth,origin=c,angle=0]{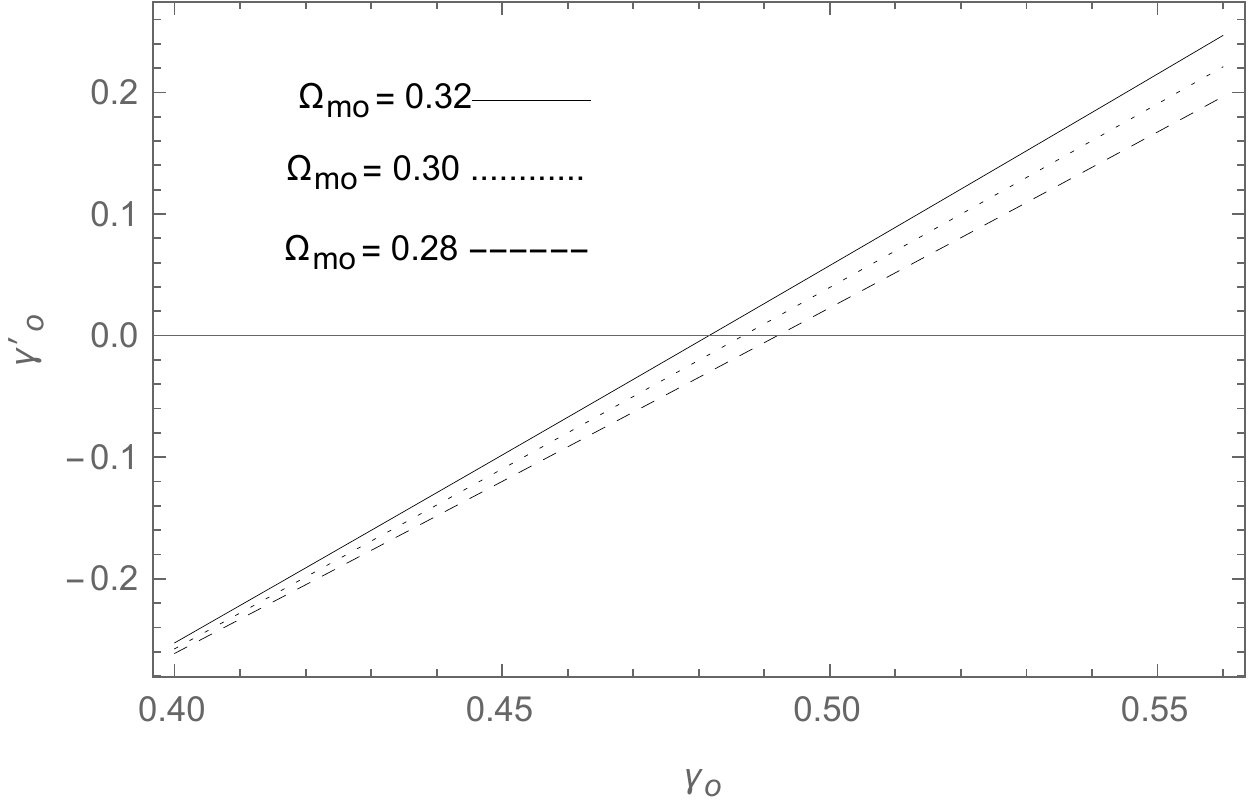}
\caption{ Plot between $\gamma^\prime_o$ and $\gamma_o$ for different values of $\Omega_{mo}$. This plot shows the small dependence of  $\Omega_{mo}$ on the  $\gamma^\prime_o$. } \label{4.3f4}
\end{figure}
We have $\gamma$ in  the $f(R)$ gravity  \citep{Narikawa:2009ux} is
\begin{eqnarray} \gamma(a) = \frac{-41 +24 \xi +\sqrt{1+24\xi}}{-70 + 48\xi} \nonumber \\ + \frac{1}{8(-143 + 24\xi)(-35 + 24\xi)^2} \nonumber \\ \times[(-41 +24 \xi +\sqrt{1+24\xi})(431 + \sqrt{1+24\xi} \nonumber \\ +24 \xi (-13 + \sqrt{1+24\xi}) -36(2+\sqrt{1+24\xi}) \nonumber \\  + 6(-2 + 3\sqrt{1+24\xi})) ](1-\Omega_m) \nonumber \\ + \mathcal{O}((1-\Omega_m)^2), \label{IV5}\end{eqnarray}
where $G_{eff}/G = \xi$.
Now we use $\Omega_{mo} \simeq 0.32 $ and $ w_{eff}$ from the equation \eqref{IIA11} for calculating the $\gamma_o $ at $z = 0$ from the equation \eqref{IV5}.  This is given as
\begin{eqnarray} \gamma_{o} = 0.547. \end{eqnarray}
Hence, we can say that our results are very close to the $\Lambda$CDM model.
After putting the value of the $G_{eff}/G_N $ from the \eqref{IVA1} in the equation \eqref{IIIA10} to calculate the value of $ \gamma_o^\prime$  for the model $f(R) = R+R^{1+\delta}/R_c^\delta$,  we obtain
\begin{eqnarray} \gamma^\prime_{o} = 0.205595. \end{eqnarray}
The large value of the $\gamma^\prime_o$ shows  clear distinction from  the dark energy in the GR background and from the $\Lambda$CDM model. The parameter $\gamma^\prime$ increases the transparency in the different types of dark energy models. The variation of $\gamma^\prime_o$ at different values of matter density $\Omega_{mo} $ with  $\delta$ in Fig.\eqref{4.3f3}, shows $\gamma^\prime_o = 0 $ if $\delta $ is around 0.3. This  means that we find the signature of the $f(R)$ in this model for $\delta>0.3$. This gives a large range of the model parameter $\delta$ in large scale structure formation and Figure \eqref{4.3f4}  also shows the large value of $\gamma^\prime_o$.

\section{\label{6} Conclusion }
In this study we conclude that the growth rate of the formation of structures is substantially influenced  by the form of $f(R)$ in the modified gravity models.  We find the value of $\gamma_o$ for the  model given by \eqref{IVC1} with  $\delta = 0.98$ and $\Omega_{mo} = 0.32$ to be $0.547$. This value is in very close agreement with the $\Lambda$CDM model. But the value of $|\gamma\prime_o| = 0.206$ which is larger than  $\Lambda$CDM value, where $\gamma\prime_o = -0.015$ in GR background. This allows us to distinguish this model from the $\Lambda$CDM model. We notice a the small dependence of $\gamma^\prime_o$ on the value of $\Omega_{mo}$ and also on the model parameter $\delta$  from  Figure \eqref{4.3f3}. In Figure \eqref{4.3f4},  we show that our model gives the value of $\gamma_o$  within the range of observational values from Dark Energy Survey (DES)/Planck/JLA/BAO \citep{LIGOScientific:2017ikf} is $\gamma_o = 0.640 \pm 0.076$  and from the Planck/TT+TE+EE+Low+lensing \citep{Planck:2018vyg} is $\gamma_o = 0.68 \pm 0.089$.

There is an another quantity, the  Newtonian gravitational constant,  which is potentially modified due to $f(R)$ model given by \eqref{IVC1}. Its modification  alters the  behaviour of the matter density perturbations.

We obtained the $w \simeq -0.998 $ in our  previous paper \citep{Sharma:2022tce}. Thus, we assumed  the de Sitter type background expansion and so  the  Hubble friction term of the \eqref{IIIA8} remains almost constant.
Growth in the asymptotic future in  our model, which is given by \eqref{IVC1},  will end because $G_{eff}/G_N \Omega_m \propto a^{-5}$ at $\delta = 1$ decreases faster than the expansion term.  On the other hand, the Starobinsky model shows the continuous growth in the matter density in the asymptotic future. At the  present accelerated epoch, our model   and the  Starobinsky model, both show an enhanced the growth of the matter density  at low redshift i.e. $z<1$ because $G_{eff}/G_N \Omega_m$ is dominating over the expansion term $(2 + \dot H/H^2)$. We can see that  $G_{eff}$ is higher than unity from the Figures \eqref{4.2f1} and \eqref{4.3f5}. Our model also supports the results \citep{Linder:2018pth} where the cosmic growth ends in asymptotic future in $f(R)$ gravity and also in $\Lambda$CDM  model in GR. Therefore, our model and $\Lambda$CDM  model have the identical behaviour of the growth of the matter density perturbations.  Indeed, the real information about the model beyond  GR is encoded in the form of $G_{eff}$, $\gamma^\prime$ and  the evolution of the cosmic growth.

\begin{center}
\textit{\underline{Acknowledgments}}
\end{center}
 Authors are thankful to  IUCAA, Pune for support  under  the  associateship programme where  most of the work was done. AKS is also thankful to Vipin Sharma, Bal Krishna Yadav,  and Vaishakh Prasad for the useful  discussions  on  various aspects of matter density perturbations in modified theory of gravity.
\bibliography{paper2}
\end{document}